\documentclass[aps,prx,twocolumn,superscriptaddress,10]{revtex4-2}

\usepackage{cmbright}
\DeclareFontShape{OT1}{cmss}{m}{it}{<->ssub*cmss/m/sl}{}

\usepackage{amsmath}
\usepackage{amssymb}
\DeclareUnicodeCharacter{2212}{\textendash}
\usepackage{graphicx}% Include figure files
\usepackage{dcolumn}% Align table columns on decimal point
\usepackage{bm}% bold math
\usepackage[version=4]{mhchem} % chemical formulas and elements
\usepackage{graphicx}
\usepackage[dvipsnames]{xcolor}
\usepackage{amsmath}
\usepackage{nicematrix}
\usepackage{textcomp, gensymb}
\usepackage[bb=boondox]{mathalfa}
\usepackage{subfiles}
\usepackage{tabularx} %fancy tables with fixed width
\usepackage{overpic}
\usepackage[linktocpage=true,
  colorlinks=true, 
  pdfborder={0 0 0}, 
  linkcolor=blue,
  citecolor=red,
  filecolor=yellow,
  urlcolor=blue,
  bookmarks,
  pdfauthor={},
]{hyperref}
\usepackage{blkarray}
\usepackage{lmodern}
\usepackage{comment}
\usepackage{wasysym}
\usepackage{pdfpages}
\makeatletter
\AtBeginDocument{\let\LS@rot\@undefined}
\makeatother

\usepackage[noindentafter]{titlesec}

\titleformat{\section}
  {\normalfont\large\bfseries\MakeUppercase}{\MakeUppercase{\thesection}}{0pt}{}
\titleformat{\subsection}
  {\normalfont\bfseries}{\thesection}{0pt}{}
\titlespacing{\section}{0cm}{0.7cm}{0.01cm}
\titlespacing{\subsection}{0cm}{0.45cm}{0cm}

\usepackage{xspace}
\usepackage{tabularx}
\usepackage{array}   
\newcolumntype{L}{>{$}l<{$}} 
\newcolumntype{R}{>{$}r<{$}} 
\newcolumntype{C}{>{$}c<{$}} 

\makeatletter
\renewcommand{\fnum@figure}{Figure~\thefigure}
\makeatother

\usepackage[capitalise]{cleveref}
\usepackage{placeins}
\crefname{figure}{Figure}{Figures}

\newcommand{\rucl}{$\alpha$-RuCl$_3$\xspace}

\begin{document}

\title{\rucl intercalated into graphite: a new three-dimensional platform for exotic quantum phases}

\author{Aleksandar Razpopov}
\email{razpopov@itp.uni-frankfurt.de}
\affiliation{Institut f\"ur Theoretische Physik, Goethe-Universit\"at, 60438 Frankfurt am Main, Germany}

\author{Shirin Mozaffari}
\email{shirinm@clemson.edu}
\affiliation{Department of Materials Sciences and Engineering, The University of Tennessee, Knoxville, TN 37996, USA}
\affiliation{Department of Physics and Astronomy, Clemson University, Clemson, SC 29634, USA}

\author{Takahiro Matsuoka}
\email{tmatsuoka@up.edu.ph}
\affiliation{Department of Materials Sciences and Engineering, The University of Tennessee, Knoxville, TN 37996, USA}
\affiliation{National Institute of Physics, University of the Philippines Diliman, Quezon, Metro Manila 1101, Philippines}

\author{Matthew Cothrine}
\affiliation{Department of Materials Sciences and Engineering, The University of Tennessee, Knoxville, TN 37996, USA}

\author{Nan Huang}
\affiliation{Department of Materials Sciences and Engineering, The University of Tennessee, Knoxville, TN 37996, USA}

\author{Miaofang Chi}
\email{chim@ornl.gov}
\affiliation{Center for Nanophase Materials Sciences, Oak Ridge National Laboratory, Oak Ridge, TN 37831, USA}
\affiliation{Department of Mechanical Engineering and Materials Science
Duke University, Durham, NC 27708, USA}

\author{Roser Valent\'i}
\email{valenti@itp.uni-frankfurt.de}
\affiliation{Institut f\"ur Theoretische Physik, Goethe-Universit\"at, 60438 Frankfurt am Main, Germany}

\author{David Mandrus}
\email{dmandrus@utk.edu}
\affiliation{Department of Materials Sciences and Engineering, The University of Tennessee, Knoxville, TN 37996, USA}

\begin{abstract}

Multilayer graphene with different stacking sequences has emerged as a powerful setting for correlated and topological phases.
In parallel, progress in graphene heterostructures with magnetic or correlated materials—most notably the Kitaev candidate \rucl—has demonstrated charge transfer, magnetic proximity effects, and interfacial reconstruction, creating new opportunities for engineered quantum systems.
Motivated by these developments, we explore a three-dimensional analogue in which \rucl layers are inserted directly into the van der Waals gaps of graphite, forming an intercalated system.
Here, we report the successful synthesis and comprehensive characterization of graphite intercalated with \rucl.
Using a combination of X-ray diffraction, quantum oscillation measurements, scanning transmission electron microscopy and first-principles electronic structure calculations, we study the structural and electronic properties of this intercalated crystals.
Our results demonstrate that graphite intercalated with \rucl offers a robust route to develop three-dimensional materials with access to novel correlated and topological states.

\end{abstract}

\date{\today}
\maketitle

\section{Introduction}

In recent years, multilayer Bernal-stacked graphene (AB) and graphene with rhombohedral (ABC) stacking have emerged as  remarkable platforms for correlated and topological quantum phenomena~\cite{de2022cascade,li2023charge,wagner2024,henck2018flat,dong2024anomalous,bernevig2025berry}. In particular, experimental studies of rhombohedral graphene  have revealed the formation of flat-band states, interaction-driven insulating phases, magnetism, and superconductivity in devices with just a few stacked layers. For example, in tetralayer or higher ABC-stacked graphene the interplay of low-density carriers, large Berry curvature, and strong Coulomb interactions has given rise to anomalous Hall and superconducting states~\cite{zhou2024layer,choi2025superconductivity,han2025signatures}. These advances underscore the power of stacking geometry, interlayer coupling, and band flattening to access exotic electronic ground states in a purely carbon-based van der Waals system.

At the same time, heterostructures coupling graphene with two-dimensional magnetic or correlated materials are rapidly expanding the playground of quantum phenomena. One prominent example is the van der Waals assembly of graphene with \rucl~\cite{mashhadi2019spin,zhou2019evidence,biswas_electronic_2019,leeb2021anomalous,rizzo2020charge}, a layered Mott insulating Kitaev material candidate~\cite{banerjee2016proximate,winter2017breakdown,winter2017models,trebst2022kitaev}.
In these \rucl/graphene heretostructures, transport experiments provide evidence for both charge transfer and magnetic proximity effects~\cite{mashhadi2019spin,zhou2019evidence, leeb2021anomalous}, while theoretical studies point to strain, enhanced Kitaev interactions, and interface-driven electronic reconstruction~\cite{biswas_electronic_2019,leeb2021anomalous}.
The observation of nanometer-scale p–n junctions in STM measurements~\cite{balgley2022} further illustrates the richness of the interfacial physics in these systems~\cite{ojeda2025}. 

In this work, we report the successful growth and characterization of graphite intercalated with \rucl. By combining synthesis, X-ray measurements, quantum oscillation studies, scanning transmission electron microscopy (STEM) and first-principles electronic structure calculations, we fully characterize the samples and introduce with them a new class of three-dimensional crystals.
\begin{figure*}
	\centering
	\begin{overpic}[width=0.95\textwidth,percent,grid=false,tics=4]{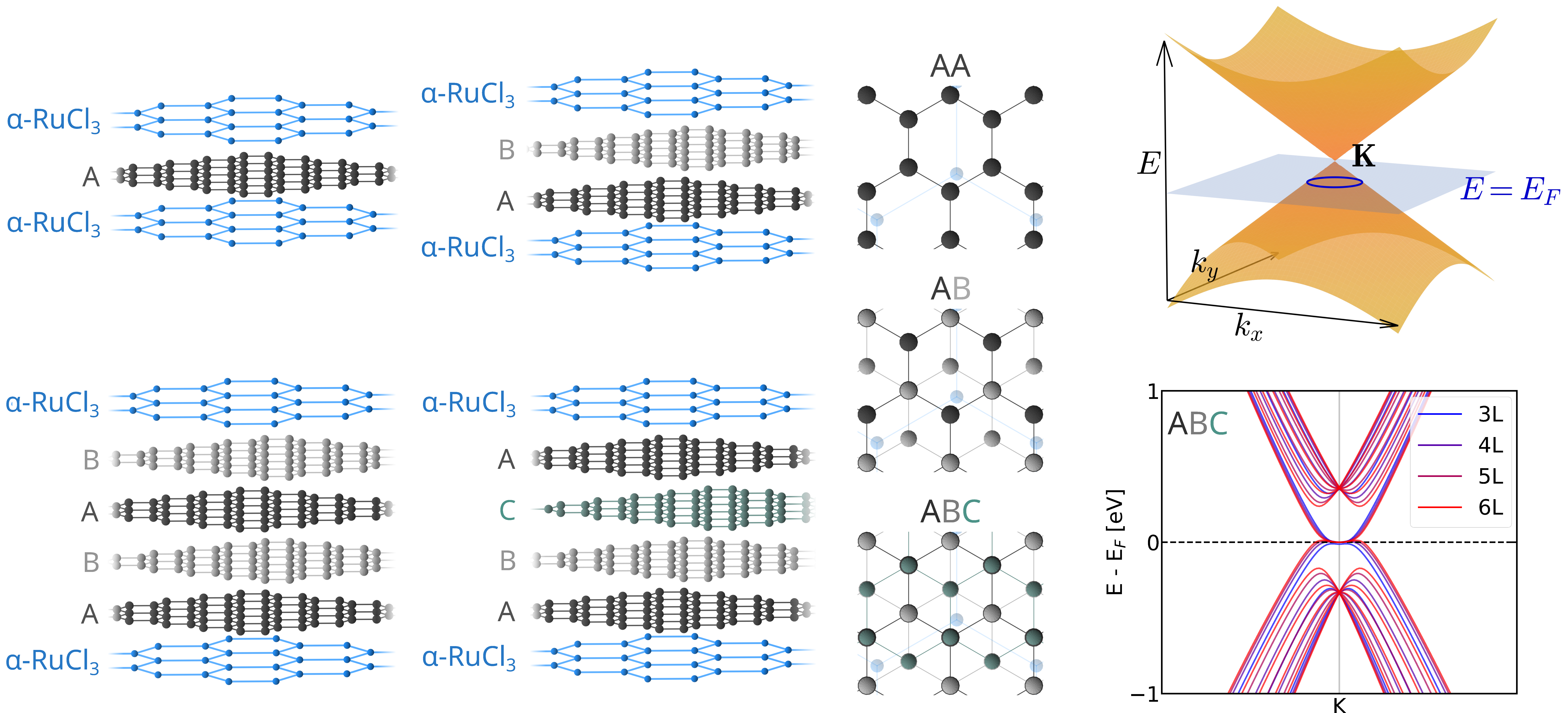}
		\put(0,43.5){\textbf{a}}
		\put(0,24){\textbf{c}}
		\put(26,43.5){\textbf{b}}
		\put(26,24){\textbf{d}}
		\put(52,43.5){\textbf{e}}
		\put(70,43.5){\textbf{f}}
	\end{overpic}
	\caption{Different intercalation stackings: \textbf{a}  A$\mid$A,  \textbf{b} AB$\mid$AB (stage 2: S2), \textbf{c} ABAB$\mid$ABAB (stage 4: S4), and \textbf{d} ABCA$\mid$ABCA (stage 4: S4). \textbf{e} Stacking top view. The \rucl layer is denoted by a blue layer, while the labels A,B, and C denote the different graphene stackings. \textbf{f} The upper panel illustrates charge transfer effects due to proximity of \rucl to graphene. The lower panel depicts the evolution of the band structure in ABC-stacked graphene.}\label{fig:stacking_types}
\end{figure*}
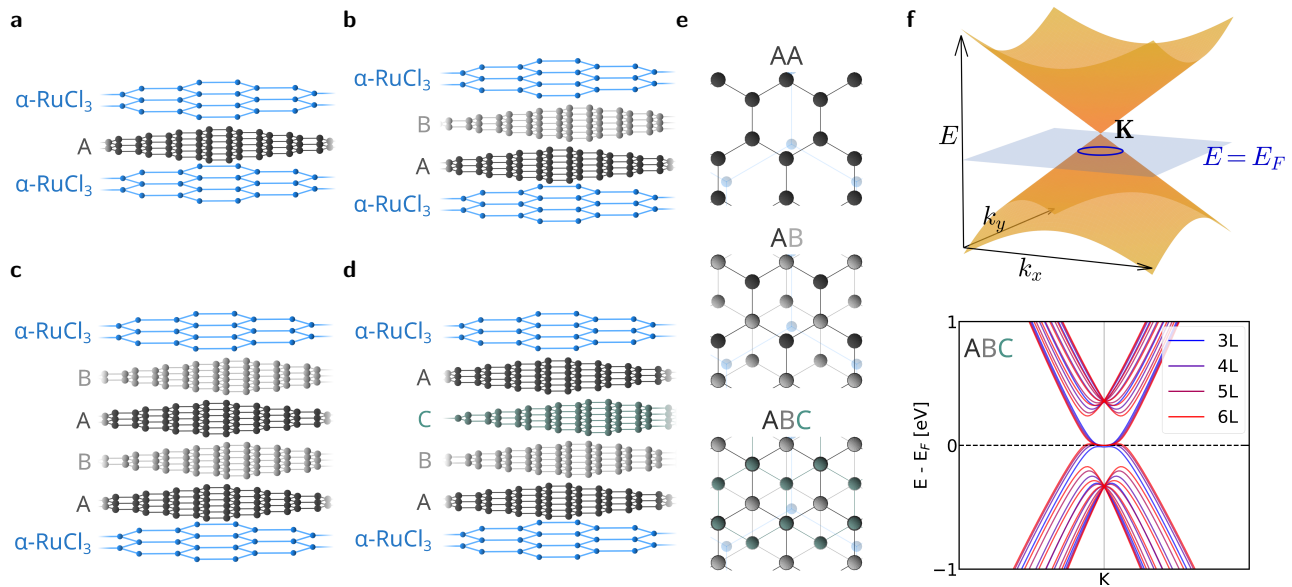

These findings open routes to engineer interfaces that combine flat-band physics, magnetism, and proximity effects, following a long tradition of magnetically intercalated graphite~\cite{dresselhaus_intercalation,chernyshev2022}.
Bringing these threads together, the prospect of graphite intercalated with \rucl (i.e., inserting layers of \rucl into the van der Waals gaps of graphite or few-layer graphene) is particularly compelling. On the one hand, the flat-band and correlation physics of rhombohedral graphene suggest that modifier layers – such as intercalants – can tune the carrier density, screening, and interlayer coupling in ways that may enhance or stabilize exotic phases. On the other hand, the graphene/\rucl heterostructure results show that \rucl is a highly effective electron acceptor and magnetically active component with possible spin-liquid behavior  in close proximity to graphene sheets. 
Intercalating \rucl into graphite therefore provides a pathway to realizing a three-dimensional analogue of these phenomena, as illustrated in~\cref{fig:stacking_types}.

\section{Results and Discussion}
\subsection{Synthesis}
\begin{figure*}[]
    \centering
    \begin{overpic}[width=1\linewidth]{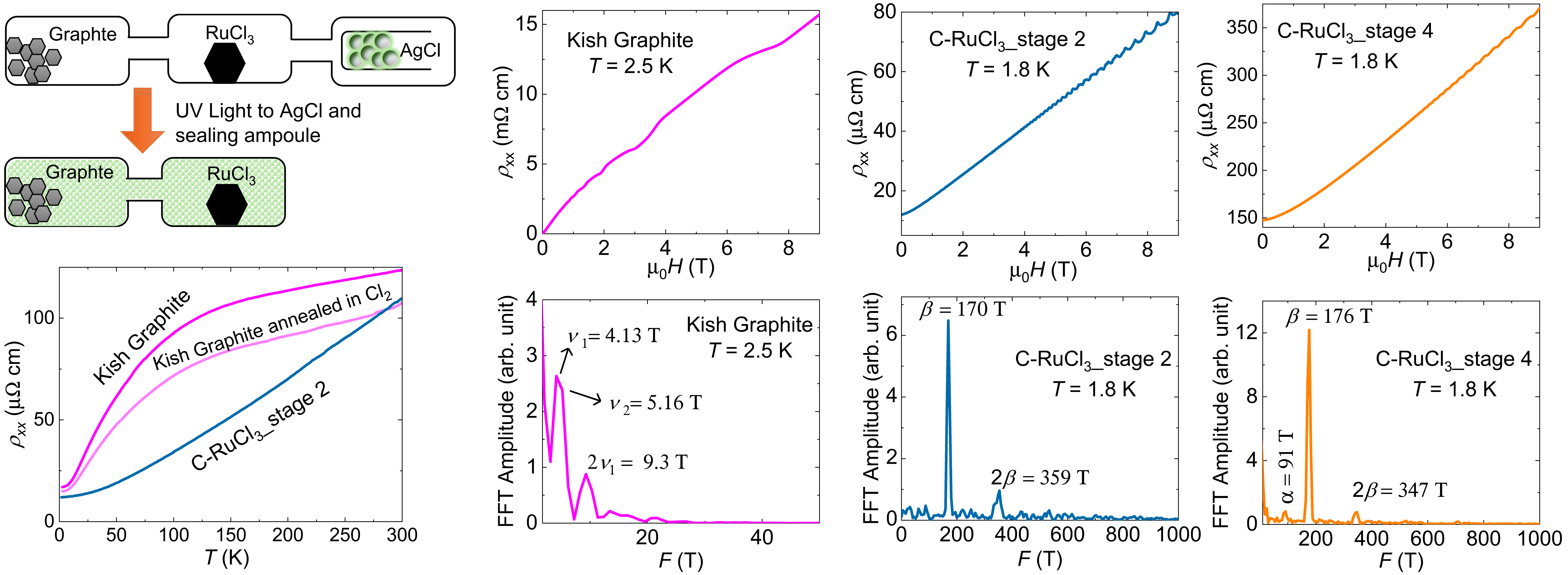}
      \put(1,37){\textbf{a}}
      \put(1,19){\textbf{b}}
      \put(32,37){\textbf{c}}
      \put(54,37){\textbf{d}}
      \put(77,37){\textbf{e}}
      \put(32,19){\textbf{f}}
      \put(54,19){\textbf{g}}
      \put(77,19){\textbf{h}}
    \end{overpic}
    \caption{\textbf{a} Schematic of two-zone vapor transport using AgCl as a chlorine source to grow C-RuCl$_3$ samples. \textbf{b} Temperature dependence of the longitudinal resistivity of Kish graphite as received and after annealing in Cl$_2$ gas, and C-RuCl$_3$ stage 2 sample. 
    \textbf{c}, \textbf{d}, and \textbf{e} Magnetic field dependence of the longitudinal resistivity $\rho_{xx}$ of Kish graphite, C-RuCl$_3$ stage 2, and C-RuCl$_3$ stage 4 samples, respectively. \textbf{f}, \textbf{g}, and \textbf{h} Fast Fourier transform (FFT) of the oscillatory component of resistivity shown in \textbf{c}, \textbf{d}, and \textbf{e} for ${\mu_0 H}$ applied out-of-plane. Peaks correspond to the extremal cross-sectional areas of the Fermi surface and are labeled as $\nu$ for Kish graphite, $\alpha$ and $\beta$ for the intercalated samples. }\label{fig:dHvA_exp}
\end{figure*}
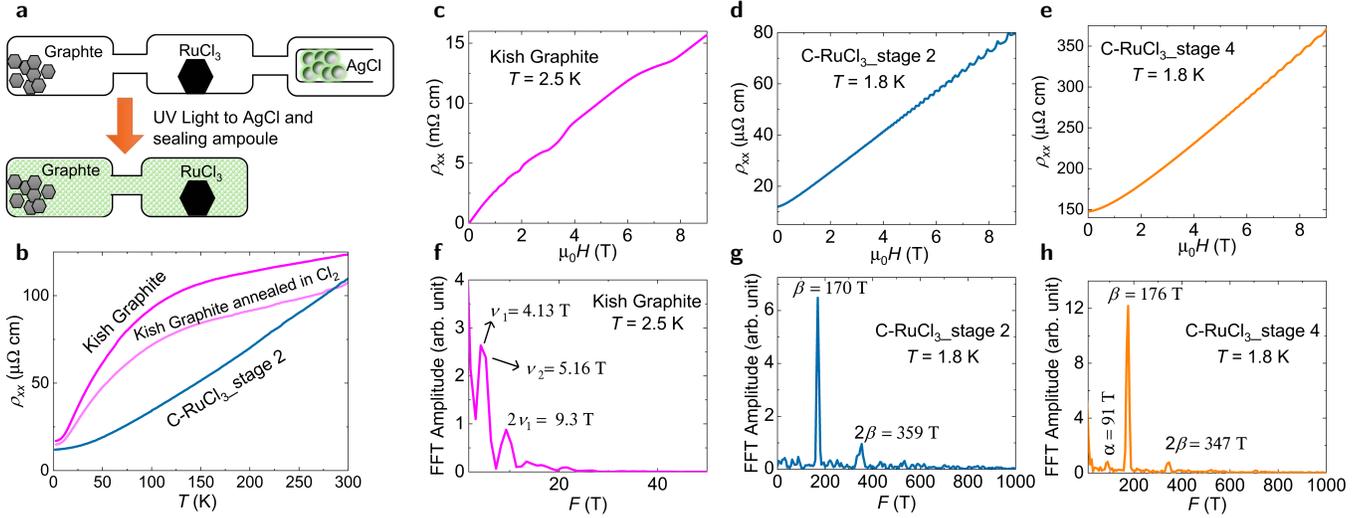

The synthesis of graphite intercalates can be performed using an intercalate in the vapor, liquid, or solid phase. Fabrication methods include solid-state, molten salt, two-zone vapor transport intercalation, and hydrothermal synthesis~\cite{dresselhaus_intercalation,synthesis_Hui,synthesis_matsumoto,synthesis_flandrois,synthesis_tzeng, synthesis_postnikov}.
Among these, two-zone vapor transport is most widely used to intercalate volatile material into the graphite. 
We grew the intercalated graphite samples via a chemical vapor transport (CVT) method.
Three starting constituents were sealed in a quartz ampule: 1) single crystals of graphite, in the form of natural graphite flakes from HQ graphene (99.98\% purity), 2) pure \rucl powder from Furuya Metals, and 3) AgCl powder from Sigma Aldrich (99.998\% purity). 
The weight ratio used to synthesize intercalated graphite was C : \rucl : AgCl  0.03g : 0.2g : 0.5g. 
The \rucl powder was pressed into a pellet to keep it together, and the AgCl powder was initially kept in an alumina crucible. 
We found that without excess Cl$_2$, \rucl would not intercalate into graphite.
To generate a Cl$_2$-rich atmosphere, Cl$_2$ gas was generated by shining UV light on the AgCl powder until it decomposed, leaving behind Ag powder.
We used powdered AgCl to increase the surface area exposed to UV light. 
Typically, the sample was irradiated with UV light for 3–5 minutes, until most of the powder turned dark brown.
The AgCl section of the tube was then sealed off, leaving the graphite flakes, \rucl pellet, and an enhanced Cl$_2$ atmosphere. This is schematically shown in~\cref{fig:dHvA_exp}\textbf{a}.
After the ampule was prepared, synthesis was carried out in a two-zone tube furnace. The graphite end was placed in the high temperature zone, which was held at 900~$^\circ$C, while the \rucl end was held at 830~$^\circ$C. After 22 hours, stage 4 \rucl intercalated into graphite, that we denote C-RuCl$_3$ was formed. Continuing the reaction for a total of 48 hours resulted in stage 2 C-RuCl$_3$. Continuing the reaction further led to mixed stage intercalation and/or disordered compounds.

\subsection{Structure analysis}
By utilizing the structural analogy to other transition metal trichlorides, different intercalation stackings shown in \cref{fig:stacking_types} are anticipated for \rucl-intercalated graphite~\cite{Dresselhaus2002,Cowley1956, Stumpp1977}.
In \cref{fig:stacking_types}, A, B and C denote different orientations of the graphene layers and "$\mid$" represents one \rucl layer.
X-ray diffraction (XRD) experiments of powdered samples resulted in two values for the c-axis and both of them show a significant expansion with respect to the c-axis of graphite (6.7 \AA $\rightarrow$ 25.4(2) \AA\: and 58.2(1) \AA), suggesting the successful intercalation of \rucl between the layers (see Figures S1\textbf{a} and \textbf{b} in the Supplemental Material.). From the structural analogy to other transition metal-trichlorides, we concluded the synthesized samples are stage 2 (S2) and stage 4 (S4) samples (\cref{fig:stacking_types}\textbf{b}, \textbf{c}, and \textbf{d}). In the S2 sample, two graphene layers are sandwiched by two \rucl layers, while in S4, four graphene layers are sandwiched by two \rucl layers.
XRD spots from single crystal (highly oriented polycrystalline) S2 and S4 samples can be indexed with hexagonal structures, with a small portion of residual pure graphite (see Figures S1\textbf{b} and \textbf{c} in the Supplemental Material). The estimated \textit{a}-axes are 29.62(3) \AA\: for S2 and 29.5(A) \AA\: for S4. The $N_1 \times M_1 \times 1$  graphene and $N_2 \times M_2 \times 1$ \rucl unit cells alternate in the crystallographic $c$-direction forming super lattices in both samples, where \textit{N$_1$}, \textit{N$_2$}, \textit{M$_1$}, \textit{M$_2$} are integers.
 
The obtained lattice constants and stackings are summarized in~\cref{tag:XRD_Exp_lattice} and~\cref{fig:stacking_types}. For the details of experimental setups and the structure analysis, see the Methods section and the Supplementary Material.

\subsection{Transport, Shubnikov-de Haas quantum oscillations and STEM measurements}
\cref{fig:dHvA_exp}\textbf{b} shows the temperature dependence of the longitudinal resistivity for three samples: pristine Kish graphite, Kish graphite annealed in Cl$_2$ gas, and graphite intercalated with $\alpha$-RuCl$_3$, denoted C-RuCl$_3$. Pristine Kish graphite was used as a reference. To isolate the effects of the synthesis procedure, we also annealed graphite in a chlorine atmosphere at 900~$^\circ$C for several days in a sealed tube without a \rucl pellet ensuring any observed changes were not solely due to the annealing environment.
It is important to note that the reported resistivity values should not be considered absolute, as the samples are flaky and irregular in shape. While we made careful efforts to estimate the sample dimensions, uncertainties remain. In particular, we found that using a blade to carve bar-shaped pieces sometimes alters the residual resistivity ratio [(RRR = R(300K)/R(2K)], likely due to mechanical damage. 

\cref{fig:dHvA_exp}\textbf{c}--\textbf{e} present the low-temperature magnetic field dependence of the resistivity for the Kish graphite, stage 2 and stage 4 intercalated samples. For the Kish graphite, Shubnikov–de Haas (SdH) quantum oscillations are observable at magnetic fields as low as 1.5~T. The intercalated samples exhibit SdH oscillations at $\mu_0H > 4$~T, and with much shorter periods. The frequency of the SdH oscillations was obtained by first subtracting a smooth polynomial background from the magnetoresistivity data to isolate the oscillatory component. The oscillatory part is periodic in the inverse of the magnetic field. A Fast Fourier Transform (FFT) was then performed on the oscillatory signal to extract the oscillation frequency. 
The frequencies we obtained for Kish graphite are $\nu_1=4.13$~T and $\nu_2=5.16$~T, with the first harmonic of $\nu_1$ also clearly observable; as shown in~\cref{fig:dHvA_exp}\textbf{f}. 
This is in good agreement with previous determinations of the fundamental SdH frequencies in graphite, including those obtained using other experimental techniques~\cite{Graphite_SdH_1,Graphite_SdH_2}. 
As shown in~\cref{fig:dHvA_exp}\textbf{g} and \textbf{h}, the frequency of the SdH oscillations in the \rucl intercalated graphite samples occurs around $\beta\approx$170--176~T. The first harmonic, shown by $2 \beta$, is also clearly observable. 
In addition to the main frequency, the C-RuCl$_3$ S4 sample shows an additional SdH oscillation frequency at $\alpha =91$T. 

To assess uniformity, we further measured quantum oscillations on a piece from the same S2 intercalated graphite crystal used for resistivity. We consistently observed the same oscillation frequency ($\beta$ peak at ~170 T) and its higher harmonics, indicating a spatially uniform electronic structure with good in-plane order and low disorder. Measurements on multiple independently prepared samples reproduced the same FFT peak, confirming uniform intercalation. We note that alternative scenarios such as disorder or mixed staging would be expected to broaden or complicate the oscillation spectrum, which is not observed. 
Hall measurements yielded a carrier density of approximately $7.8\times10^{20}$ cm$^{-3}$, as extracted from the slope of the transverse resistivity.
Finally, we also performed STEM measurements  to further investigate the structural characteristics of the system. The STEM images (see Supplementary Information) reveal an alternating contrast between adjacent layers, which is consistent with the presence of an intercalated structure. While a fully resolved layer-by-layer intercalation pattern is not observed throughout the entire sample, the observed contrast modulation provides qualitative evidence supporting successful intercalation.

\subsection{Structure optimization and charge transfer}

\begin{figure*}
	\centering
	\begin{overpic}[width=0.95\textwidth,percent,grid=false,tics=2]{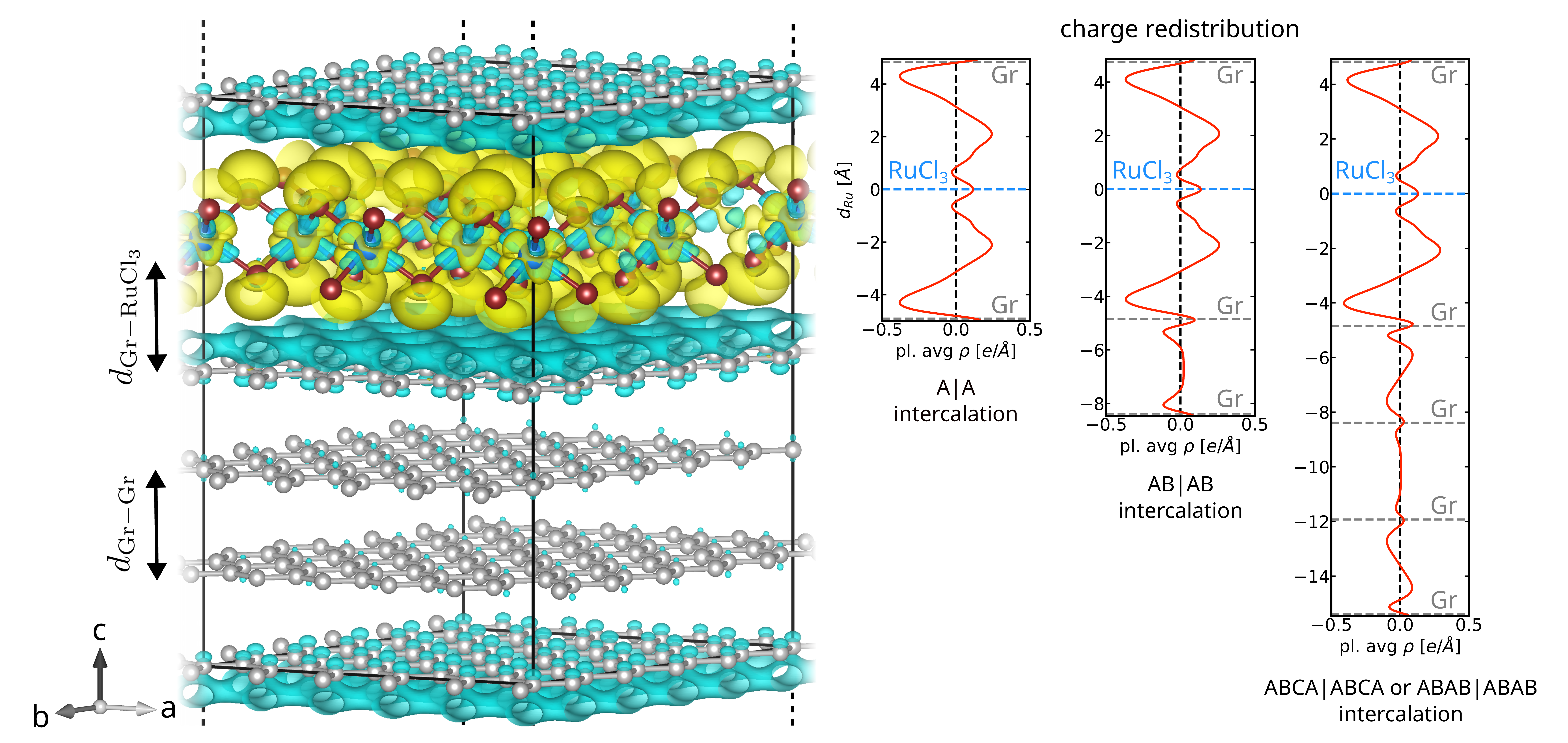}
		\put(0,47){\textbf{a}}
		\put(53,47){\textbf{b}}
	\end{overpic}
	\caption{\textbf{a} Representative ABCA$\mid$ABCA crystal structure of C-RuCl$_3$ used in the DFT calculations. The blue and yellow iso-surfaces show the charge depletion and accumulation, respectively. $d_{\rm Gr-Gr}$ denotes the graphene-graphene layer distance, and $d_{\rm Gr-RuCl_3}$ the distance \rucl-graphene layer distance. \textbf{b} The planar-averaged charge density redistribution along the c-axis for each geometry as a function of the Ru layer distance, $d_{\rm Ru}$.  The reference point is the position of the Ru layer indicated by the dashed blue line, the grey dashed lines indicate the position of the graphene layers. The four-layer intercalations in the ABAB$\mid$ABAB and ABCA$\mid$ABCA stacking are represented by one plot. Crystal structure plots are generated with VESTA~\cite{VESTA}.}\label{fig:crystal_structure_charge_transfer}
\end{figure*}

In order to have a full characterization of the grown crystal structures and extract their electronic properties, we perform {\it ab initio}-based density functional theory (DFT) calculations (see Methods section). 
Given the considerable size of the experimental unit cells reported in \cref{tag:XRD_Exp_lattice}, carrying out calculations directly on these cells is computationally prohibitive.
Therefore, we employ reduced (simplified) unit cells in which a tensile strain of $\approx$~4\% is applied to the \rucl layer to keep the overall cell size manageable, while leaving the geometry of the graphene layers unchanged.
As shown below, this simplification does not significantly affect the relevant physical observables in the transport measurements.

To verify this, we benchmark the strained structure against the unstrained one.
We consider an unstrained C-RuCl$_3$ structure of a 5$\times$5 \rucl layer intercalated into a single layer of 12$\times$12 graphene in the A$\mid$A (\cref{fig:stacking_types}\textbf{a}) stacking and a strained unit cell consisting of a 2$\times$2 \rucl layer intercalated into a single 5$\times$5 graphene layer with the same stacking type.
Both unstrained and strained structures are relaxed by keeping the strong C-C bond in graphene fixed while the \rucl layer is fully relaxed.

We compute the electronic properties of both type of structures within DFT and compare them with those of pristine single-layer graphene and bulk $\alpha$-RuCl$_3$. By aligning the electronic bands of the unstrained C-RuCl$_3$ with those of the pristine materials, we observe no significant hybridization between the graphene and \rucl layers; instead, mostly charge transfer takes place (see Figures S3 and S5 in the Supplementary Material), in agreement with previous studies on similar heterostructures~\cite{biswas_electronic_2019,leeb2021anomalous,rossi2023direct,razpopov2024ab}.
Because the hybridization is effectively very weak, the applied strain in the simplified unit cell impacts mostly the \rucl electronic structure, leaving the graphene bands unchanged except for a band shift due to charge transfer (see below).
Using the graphene-\rucl separation $d_{\rm Gr-RuCl_3}$ suggested by X-ray data, we find nearly identical charge depletion in both, strained and unstrained systems. We also tested different relative orientations and found that they produce no significant changes in the charge transfer (see Figure S3 in Supplementary Material).

Based on (i) the absence of strong hybridization between the graphene and \rucl layers, (ii) the fact that strain in the \rucl layer does not appreciably modify the charge transfer, and (iii) the charge transfer is rather insensitive to the relative orientation of the layers, we conclude that the electronic  bands can be reliably obtained using a reduced unit cell with a tensile-strained \rucl layer.

Following this approach, we model the experimental structures determined by XRD by constructing four geometries containing a tensile-strained 2$\times$2 \rucl layer intercalated with a 5$\times$5 graphene supercell in the stacking arrangements 
  A$\mid$A, AB$\mid$AB, ABAB$\mid$ABAB and  ABCA$\mid$ABCA as shown in~\cref{fig:stacking_types}.
The theoretical structures  AB$\mid$AB correspond to the S2 sample, and  ABAB$\mid$ABAB and ABCA$\mid$ABCA to the S4 sample.
Here, to reduce computational expense, we replicate the graphene stacking sequence after each \rucl layer, simplifying the original stacking 
from A$\mid$AB$\mid$B to AB$\mid$AB, and AB$\mid$BCAB$\mid$BC (AB$\mid$BABA$\mid$AB) to ABCA$\mid$ABCA (ABAB$\mid$ABAB). 
This simplification is justified because the charge transfer is insensitive to the relative orientation between \rucl and graphene, and the graphene sections are well separated by the intervening \rucl layer.

Our Bader analysis of the charge redistribution (\cref{fig:crystal_structure_charge_transfer}) shows that the graphene sheets donate charge to the \rucl sheet, and the charge mostly accumulates at the interface
(\cref{fig:crystal_structure_charge_transfer}\textbf{a}). Furthermore, the total charge accumulated on the \rucl layer increases with the number of intercalated graphene layers  (see Figure S4 in the Supplementary Material for details).
A similar trend has recently been reported in multilayer graphene heterostructures~\cite{falorsi2025interlayer}. The magnitude of the charge accumulation is consistent with previous studies of related \rucl/graphene systems~\cite{biswas_electronic_2019,balgley2022,razpopov2024ab}.
When comparing the charge transfer between the different ABAB$\mid$ABAB and ABCA$\mid$ABCA stacking configurations, we find no significant differences.

In~\cref{fig:crystal_structure_charge_transfer}\textbf{b}, we present the planar-averaged charge density along the c direction.
In the  A$\mid$A intercalation, every graphene layer exhibits the same charge depletion because of the periodicity of the unit cell. This also holds for the AB$\mid$AB intercalation, as each layer faces a \rucl surface.
In contrast, for the ABAB$\mid$ABAB or ABCA$\mid$ABCA configurations, charge depletion is concentrated primarily on the interface graphene layers, while the inner graphene layers remain screened.

\subsection{Electronic properties}

To understand the origin of the measured quantum oscillations, we estimate the corresponding frequencies using the Onsager relation~\cite{onsager1952interpretation}, $F = \frac{\hbar}{2\pi e} S,$ which links the frequencies $F$ to the extremal Fermi surface cross-sectional area $S$. These are compared with the experimental data in~\cref{fig:dHvA_exp}\textbf{g} and \textbf{h}. Since the magnetic field in the experiments is applied along the crystal $c$-axis, we consider Fermi surface cross sections perpendicular to  $k_z$.
Electronic structures and corresponding Fermi surfaces are calculated for each intercalated crystal.

\begin{figure}
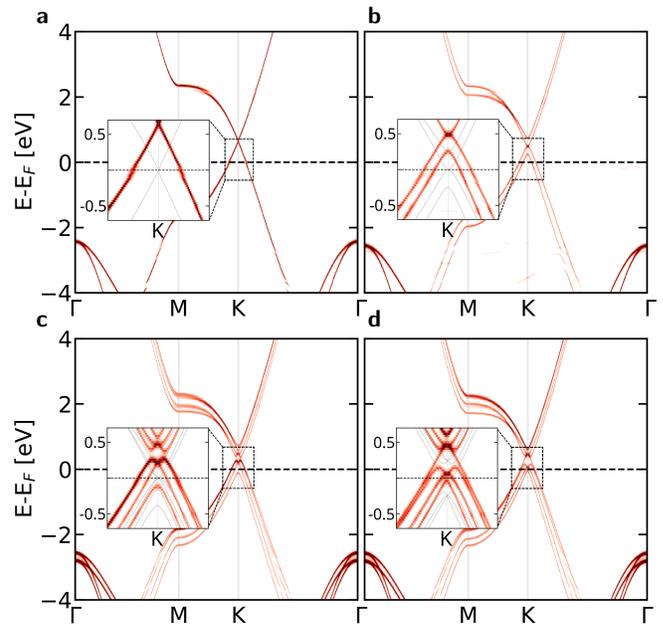

    \centering
    \begin{overpic}[width=1.0\columnwidth,percent,grid=false,tics=5]{band_structure_graphene_unfolded_withpristine_v3_arxiv_300dpi_white.png}
    \put(4,94){\textbf{a}}
    \put(55,94){\textbf{b}}
    \put(4,47.3){\textbf{c}}
    \put(55,47.3){\textbf{d}}
    \end{overpic}
    \caption{Unfolded electronic band structure of the C-RuCl$_3$ graphene layers in red for the four different intercalated geometries obtained via non-spin polarized GGA calculation, \textbf{a} A$\mid$A, \textbf{b} AB$\mid$AB, \textbf{c} ABAB$\mid$ABAB, and \textbf{d} ABCA$\mid$ABCA stacking. Each inner panel shows a zoomed region around the K point. Light gray solid lines display the bands of pristine (multi-layered) graphene.}\label{fig:bandstructure_unfold}
\end{figure}

The charge redistribution at the interface between \rucl and the adjacent graphene layer is mostly localized at the Cl atoms as shown in~\cref{fig:crystal_structure_charge_transfer}\textbf{a}
with \rucl remaining rather insulating and Ru atoms keeping their magnetic nature~\cite{zheng2024incommensurate}. 
Therefore, only the graphene layers are expected to contribute to the observed quantum oscillations. 
We compute the graphene layers electronic properties directly from the intercalated structures using non-spin-polarized GGA.
The band structures are unfolded and projected onto graphene states, then compared to pristine, undoped graphene (\cref{fig:bandstructure_unfold}). Note the formation of a flat band with the number of stacked graphene layers, as has been observed in rombohedral graphene stacking~\cite{aoki2007dependence,johansson2014multiple,yagi2018low,henck2018flat,zhang2025layer}.
In all intercalated cases, charge transfer induces hole pockets around the \textbf{K} points.

In the A$\mid$A (\cref{fig:bandstructure_unfold}\textbf{a}) and AB$\mid$AB(\cref{fig:bandstructure_unfold}\textbf{b}) intercalations, the pristine graphene bands are preserved with a global Fermi energy shift of 0.68 eV and 0.48 eV, respectively, due to equivalent hole doping of the graphene layers (\cref{fig:crystal_structure_charge_transfer}\textbf{b}). The smaller shift in the bilayer AB$\mid$AB case arises because the transferred charge is distributed across two bands, whereas in the single-layer A$\mid$A case only one band contributes to the depletion.
In the four-layer ABAB$\mid$ABAB and ABCA$\mid$ABCA stackings, the energy shift is non-uniform, reflecting the layer-dependent charge redistribution (\cref{fig:bandstructure_unfold}\textbf{c}, \textbf{d}).
The pristine four-layer structures show four energy-separated bands, in agreement with previous studies~\cite{aoki2007dependence,johansson2014multiple,yagi2018low,henck2018flat,zhang2025layer}, while in the intercalated systems, two of these bands overlap or nearly overlap due to doping.
The band geometry around  \textbf{K} also depends on the stacking, directly influencing the Fermi surface.

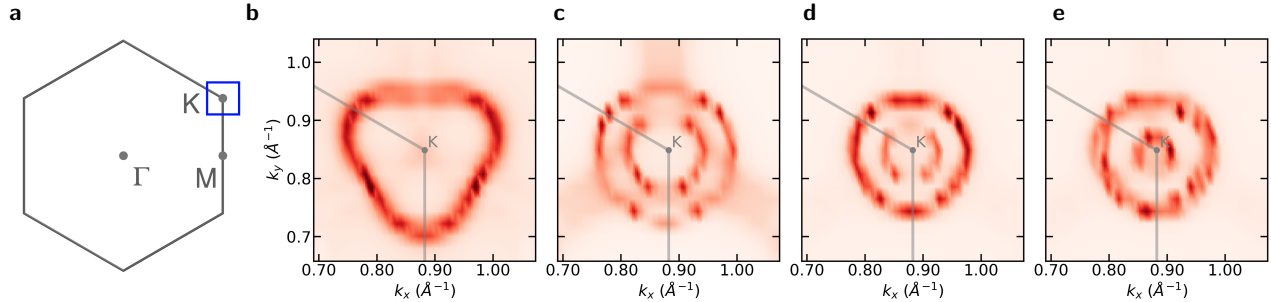
\begin{figure*}[]
    \centering
    \begin{overpic}[width=0.95\textwidth,percent,grid=false,tics=6]{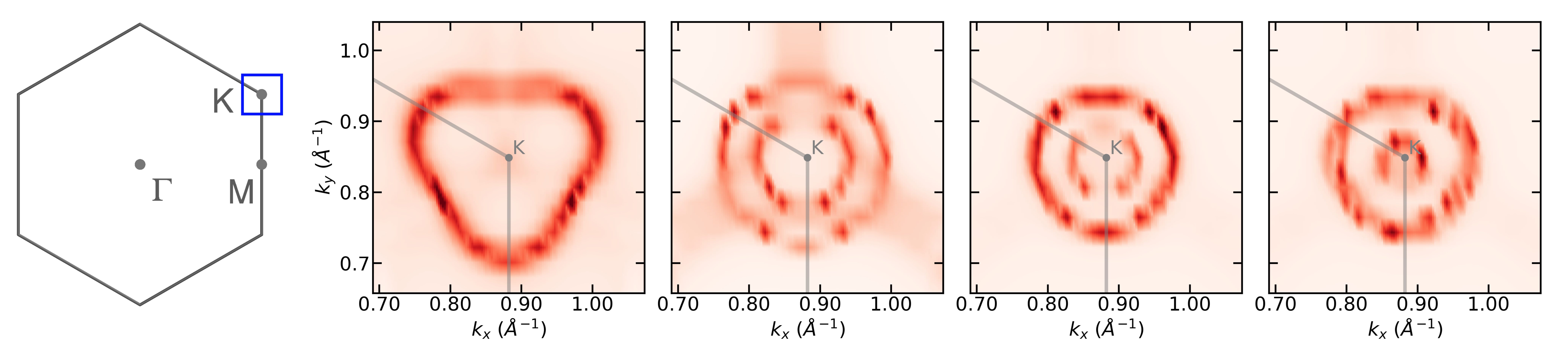}
       \put(0,22){\textbf{a}}
       \put(18.5,22){\textbf{b}}
       \put(42.5,22){\textbf{c}}
       \put(62,22){\textbf{d}}
       \put(81.5,22){\textbf{e}}
    \end{overpic}
    \caption{Unfolded Fermi surface of the graphene layers C-RuCl$_3$ structure calculated via a non-spin polarized GGA calculation. \textbf{a} Full Brillouin zone. The blue region indicates the calculated unfolded region around the K point. \textbf{b}-\textbf{e} Fermi surfaces for \textbf{b} A$\mid$A, \textbf{c} AB$\mid$AB, \textbf{d} ABAB$\mid$ABAB, and \textbf{e} ABCA$\mid$ABCA stackings.}\label{fig:fermi_surface}
\end{figure*}

The Fermi surfaces are computed by unfolding the bands and projecting onto graphene states, focusing on the region around \textbf{K} where the bands cross the Fermi level (\cref{fig:fermi_surface}).
The A$\mid$A intercalation yields a single hole pocket, consistent with prior experimental observations~\cite{mashhadi2019spin}.
AB$\mid$AB produces two pockets of different sizes, while  ABAB$\mid$ABAB and ABCA$\mid$ABCA exhibit three hole pockets.
In the ABAB$\mid$ABAB, the three pockets are distinct, with the two outer bands overlapping, forming a larger pocket plus a smaller one.
In the ABCA$\mid$ABCA, the three pockets are distinct, with the two outer bands nearly overlapping.
Thus, the stacking sequence controls the number and size of Fermi hole pockets. We compute their areas and the corresponding quantum oscillation frequencies using the Onsager equation, and compare the calculated quantum oscillation frequencies with the experimentally measured ones ($\alpha \approx91$~T, $\beta \approx 170-176$~T).

Single-layer A$\mid$A intercalation produces a single high-frequency pocket ($\approx$ 511 T), inconsistent with both quantum oscillations experiments and XRD, discarding this as the stacking pattern.
AB$\mid$AB (S2-like) yields two pockets (131 T, 444 T), while four-layer ABAB$\mid$ABAB and ABCA$\mid$ABCA (S4-like) produce two and three pockets (70T, 296 T, and 28T, 311 T, 330 T), respectively.
Although the agreement improves, discrepancies with the experiment remain. Examining their possible origin, we find that the Fermi-surface areas are highly sensitive to the interlayer distances ($d_{\rm Gr-Gr}$, $d_{\rm{Gr-RuCl_3}}$) and the amount of charge transfer. Modest Fermi-level shifts ($\approx$ 0.1 eV) -- consistent with plausible variations in interlayer spacing arising from, for example, stacking faults not included in our calculations -- bring the computed frequencies into good agreement with experiment.
A Fermi level shift of 0.1 eV yields two frequencies of 27~T and 183~T for the ABAB$\mid$ABAB stacking, and 126~T and 214~T  for the ABCA$\mid$ABCA stacking.
Further slight Fermi level shifts push the inner hole pocket below the Fermi energy, in which case no corresponding quantum oscillations can be observed.  
This behavior is consistent with the sample-dependent absence of the $\alpha$ frequency in the quantum oscillations measurements. 

From this theoretical analysis, we conclude that the quantum
oscillations response is governed by the stacking sequence, charge distribution, and structural variability. Our analysis shows that only a limited range of physically reasonable parameters, including charge transfer, reproduces the observed oscillation frequencies.
Stacking-fault effects may further influence the observations but remain beyond our current computational resolution. 

Finally, we remark that previous studies of \rucl /graphene heterostructures~\cite{biswas_electronic_2019,leeb2021anomalous,rossi2023direct,razpopov2024ab} have reported indications of tensile strain in the \rucl layers. While this effect does not appear to significantly affect the transport properties, it may influence the magnetic behavior of the system, which remains an important direction for future investigation.

\section{Conclusions}
In summary, we have successfully synthesized graphite intercalated with \rucl for the first time. 
By performing X-ray diffraction experiments we determined the crystal structure and showed that  several different stackings can be achieved during the growth process.
To characterize the properties of the system further, we performed transport experiments,  Shubnikov–de Haas (SdH) quantum oscillations measurements and Hall effect.
Compared to pristine graphite, which exhibits only small SdH frequencies, the intercalated samples exhibit much higher frequencies of approximately 91T and 170 T, indicating substantial modifications of the Fermi surface.

Examining the different stackings identified by the XRD experiments by carrying out DFT calculations we find no strong hybridization between the graphite and \rucl layers.
Instead, our theoretical results reveal significant charge transfer from the graphite layers to the \rucl layers which can be controlled by modifying the number of intercalated graphite layers between successive \rucl sheets.
The intercalation-dependent charge transfer strongly influences both the electronic band structure and Fermi surface of the crystal.
These changes lead to a modified Fermi surface, consistent with the experimental SdH frequencies.

Our results provide an initial step toward 3D structures combining the flat-band, strongly correlated, and topologically non-trivial physics of rhombohedral graphene with the unconventional magnetism, charge-transfer effects, and interfacial phenomena of graphene/\rucl heterostructures, subject to further spectroscopic and magnetic validation.

\section{Methods}

\subsection{Structure analysis:} We conducted XRD experiments for single crystal (nearly single crystal) samples in BL10XU/SPring-8 (X-ray wavelength: 0.4304 \AA, X-ray beam size: 20 $\mu$m in diameter) and powder samples using an X-ray diffractometer (Bruker D2 Phaser) at the Institute for Advanced Materials and Manufacturing at UTK (X-ray wavelength 1.542 \AA (Cu-K$\alpha$)).
We collected the XRD from a single crystal in a forward-scattering geometry with a 2-dimensional area detector (Varex Imaging XRD1611 CP3 Flat panel detector) while oscillating the sample $\pm$20 deg. XRD of the powder was challenging because grinding in a mortar could not crush the samples into fine powder.
Therefore, we gathered many single crystals in the sample holder of the diffractometer to produce a powder-like condition.
Thus, the obtained powder XRD is highly oriented.
The typical XRD images for the S2 and S4 single-crystal samples are in Figures S1\textbf{c} and \textbf{d} in the Supplemental Material. The XRD patterns for powder samples are in Figures S1\textbf{a} and \textbf{b} in the Supplemental Material.\\

First, we index the powder XRD peaks with \textit{hkl} ($h=k = 0$) and the calculated lattice constant \textit{c} are 25.4 \AA\: for the S2 and 58.2 \AA\: for the S4 samples, respectively. Those values are much larger than the typical values 6.7 \AA\: of pure graphite, suggesting intercalation by $\alpha$-RuCl$_3$. We estimated the number of RuCl$_3$ layers and figured out the stacking sequence of the graphene and $\alpha$-RuCl$_3$ layers in a unit cell based on the analogy with FeCl$_3$, CrCl$_3$, and AlCl$_3$,~\cite{Dresselhaus2002,Cowley1956,Stumpp1977} We adopted the following assumptions.
\begin{enumerate}
\item RuCl$_3$ crystallizes in a solid composed of Cl$_3$Ru$_2$Cl$_3$ and is incommensurate with the graphite host.
\item The graphene and RuCl$_3$ layers form a supercell constructed by combining $n \times m$ unit cells of RuCl$_3$ and $N \times M$ of graphene. Here, \textit{n}, \textit{m}, \textit{N}, and \textit{M} are integers.
\item The inter-layer distance between adjacent graphenes is about 3.5 $\sim$ 4 \AA, and the distance between two graphene layers that sandwich a RuCl$_3$ layer is about 9 $\sim$ 10 \AA.
\item Two adjacent graphene layers are stacked as AB, AC, or BC (Natural graphite has ABAB or ABCABC stacking). Two graphene layers that sandwich RuCl$_3$ in between stack AA, BB, or CC (all three are equivalent).
\end{enumerate}
Figures S1\textbf{a} and \textbf{b} in the Supplemental Material show the \textit{hkl} indexing for the single crystal samples. Note that all the observed peaks are for \textit{hkl} (\textit{l}$=$0) due to the large \textit{c}-axis and limited sample oscillation angle during the XRD measurements.
Based on the indexing results for powder and single crystal samples and the assumptions mentioned above, we conclude the crystal structure for the sample as shown in the~\cref{tag:XRD_Exp_lattice}and ~\cref{fig:stacking_types}\textbf{b}, \textbf{c} and \textbf{d}.

\begin{table}
\centering
\begin{tabular}{c|c|c|c}
\hline
sample &$a$ [\AA] & $c$ [\AA] & Stacking\\ 
\hline \hline
stage 2 &29.62(3) & 25.4(2) &  A$\mid$AB$\mid$B \\
\hline
stage 4 &29.5(4)& 58.2(1) & AB$\mid$BCAB$\mid$BC or AB$\mid$BABA$\mid$AB \\
\hline
\end{tabular}
\caption{Experimentally obtained lattice constants for c-RuCl$_3$ and the suggested stacking of layers. The vertical line $\mid$ indicates the $\alpha$-Cul$_3$ layer.}\label{tag:XRD_Exp_lattice}
\end{table}

\subsection{Quantum oscillations:}
Magnetotransport experiments were performed in a physical property measurement system (PPMS-Quantum Design) under magnetic fields up to 9~T and temperatures as low as 1.8~K using a conventional four-terminal method. The measurement was performed by passing the current in the \textit{ab}-plane and applying the magnetic field along the \textit{c}-axis of the crystal.  The measured longitudinal and transverse resistivities were field-symmetrized and antisymmetrized, respectively, to correct the effect of contact misalignment. 

\subsection{STEM measurements:}
Cryogenic STEM imaging was performed using a JEOL NEOARM transmission electron microscope operated at 80 kV with a convergence semi-angle of 26.9 mrad and a HAADF collection angle ranging from 68 to 280 mrad.
A Mel-Build double-tilt LN2 cooling cryogenic holder was used to cool the specimen to 100 K in order to minimize electron-beam-induced damage.
Each image was acquired as a stack of 15 frames with dwell times of 0.25-1 µs per pixel.
The image stacks were aligned using cross-correlation and subsequently summed to reduce image instability caused by sample drift and to improve the signal-to-noise ratio.

\subsection{Computational details:}
We perform density functional theory (DFT) calculations as implemented in VASP~\cite{Kresse_VASP}~simulation package version 6.3.0.
In all calculations we apply the Generalized Gradient Approximation (GGA)~\cite{GGA} as exchange-correlation functional.
We use the following pseudopotentials in the projector augmented wave method (PAW)~\cite{blochl_projector_1994,kresse_ultrasoft_1999} Ru\_pv, Cl and C provided by the VASP package.
In all calculations the basis set plane-wave cut-off for the expansion is set to 500~eV.
In all crystal structures we relax the \rucl layer and keep the graphene layers geometry fixed.
The graphene-\rucl~layer distance is kept fixed at approximately the average experimental value of $d_{\rm{Gr-RuCl_3}}=3.60$~\AA, the adjacent graphene sheets at $d_{\rm Gr-Gr}=3.54$~\AA, and the C-C distance in each graphene layer is $\approx1.42$~\AA, consistent with previous reports~\cite{Chen2011_tight_binding_graphene}.
The relaxations are performed via the $\Gamma$-point version of VASP with a convergence criterion of the forces for each atom in the \rucl~layer in each direction of 0.005 eV/\AA~assuming a non-magnetic state.
For the crystal cell optimization we include van der Waals corrections via the DFT-D3 method of Grimme~\cite{grimme2010consistent,grimme2011effect}.

Further electronic properties are calculated on a 6$\times$6$\times$2 k-mesh.
The supercell electronic band structures are unfolded into the primitive Brillouin zone (BZ) by using the KPROJ software~\cite{chen2025kproj}.
The Fermi surfaces are generated from sequences of unfolded electron band structures.
The charge transfer was estimated via the Bader analysis~\cite{HENKELMAN2006354}, and k-mesh convergence has been checked.
The spatial resolved charge transfer was calculated via the VASPKIT package~\cite{VASPKIT}.
The quantum oscillation frequencies are calculated from the unfolded graphene layers Fermi surfaces.
Due to the smeared Fermi surfaces as a consequence of the unfolding, the Fermi pocket boundaries are taken from the highest intensity points and connected to form a closed area.
Therefore, all reported frequencies inherit a small systematic error which does not change the conclusions of the study.

Additionally, as a cross-check we compute the quantum frequencies for the A$\mid$A structure via the C++ program dhva~\cite{backes2014electronic,numpy2024}, which implements the dHvA frequency extraction algorithm introduced by Rourke and Julian~\cite{julian2012numerical}. Calculations are performed on a supercell grid comprising $6.4\times10^{7}$ k-points distributed over 64 unit cells. Higher k-point density is obtained through tricubic interpolation~\cite{lekien2005tricubic}. The magnetic field is applied in the (001) direction.

\medskip
\textbf{Acknowledgements} \par 
A.R and R.V. thank the Deutsche
Forschungsgemeinschaft (DFG, German Research Foundation) for funding through the TRR 288 - 422213477 (project A05, B05) and  gratefully acknowledge the computing time provided to them on the Goethe-HLR cluster at the Frankfurt Center for Scientific Computing. A.R. thanks Daniel Guterding for support of the hdva code (Github).
S.~M. and D.~M. acknowledge the support from the Gordon and Betty Moore Foundation's EPiQS initiative, Grant GBMF9069 and the support from AFOSR MURI Grant No. FA9550-20-1-0322.

X-ray diffraction was performed at the Institute for Advanced Materials $\&$ Manufacturing (IAMM) Diffraction facility, located at the University of Tennessee, Knoxville, and at BL10XU/SPring-8 (Proposal number 2024B1285).

Electron microscopy was supported by the U.S. Department of Energy, Office of Basic Energy Sciences, Division of Materials Sciences and Engineering, and was performed at the Center for Nanophase Materials Sciences (CNMS), a U.S. DOE Office of Science User Facility at Oak Ridge National Laboratory (ORNL). 

\medskip
\textbf{Author Contributions} \par
N.H. and M.C synthesized the crystals.
S.M. performed the transport and quantum oscillation measurements.
T.M. and M.C. performed the X-ray experiments. 
M. C. performed the cross-sectional STEM measurements. 
A.R. performed the \textit{ab initio} simulations.
R.V. and D.M. conceived and supervised the project.
All authors contributed to the discussions and writing of the manuscript.

\medskip
\textbf{Corresponding authors}
Correspondence to Aleksandar Razpopov (razpopov@itp.uni-frankfurt.de) and Roser Valent\'i (valenti@itp.uni-frankfurt.de).

\medskip
\textbf{Conflict of interest}  \par 
The authors declare no conflict of interest.

\medskip
\textbf{Data availability statement} \par 
The datasets generated during the current study are avail-
able upon reasonable request.
The authors declare no conflict of interest.

\newpage
\clearpage


\section*{Supplementary material}

# $\alpha$-RuCl$_3$ intercalated into graphite: a new three-dimensional platform for exotic quantum phases - Supplementary Information -

Aleksandar Razpopov,[1, *] Shirin Mozaffari,[2, 3, †] Takahiro Matsuoka,[2, 4, ‡]
Matthew Cothrine,[2] Nan Huang,[2] Roser Valentí,[1, §] and David Mandrus[2, ¶]

[1]*Institut für Theoretische Physik, Goethe-Universität, 60438 Frankfurt am Main, Germany*
[2]*Department of Materials Sciences and Engineering,*
*The University of Tennessee, Knoxville, TN 37996, USA*
[3]*Department of Physics and Astronomy, Clemson University, Clemson, SC 29634, USA*
[4]*National Institute of Physics, University of the Philippines Diliman, Quezon, Metro Manila 1101, Philippines*

## X-RAY DIFFRACTION ANALYSIS

In Figure S1 we show the XRD images and the integrated XRD profiles for stage 2 and stage 4 C-RuCl$_3$ samples.

## SUPERCELLS CALCULATIONS WITH DIFFERENT ORIENTATIONS

We analyze the influence of relative orientation between the $\alpha$-RuCl$_3$ and the Graphene layers on the charge transfer. Therefore, we calculate the electron properties of one intercalated 5×5×1 $\alpha$-RuCl$_3$ layer sandwiched between single 12×12×1 Graphene layers alternating in the crystallographic c-direction, as shown in Figure S2**a**.

We consider two different relative orientations A and B. In both initial geometries the graphene-graphene layer distance ($d_{\mathrm{Gr-Gr}}$) is 10.241 Å and the distance between the $\alpha$-RuCl$_3$ and the graphene layer ($d_{\mathrm{Gr-RuCl_3}}$) is 3.765 Å. The C-C distance ($d_{\mathrm{C-C}}$) within the graphene layer is 1.434 Å, the Ru-Ru distance in the $\alpha$-RuCl$_3$ layer is 3.443/3.441 Å. This corresponds to a tensile strain of $\approx 1\%$ compared to the relaxed C2/m structure of $\alpha$-RuCl$_3$ [1]. The structures are fully optimized via non-spin polarized GGA calculation using the VASP package [2].

For the relaxed structures we calculate the non-spin polarized electronic band structures and density of states. The results for both orientations A and B are shown in Figure S3. Due to a significant charge transfer from graphene to $\alpha$-RuCl$_3$, the Dirac cone of graphene is shifted $\approx 0.7$ eV above the Fermi surface. We note that due to the super-cell choice of the graphene layer, the K point is folded at the $\Gamma$-point, see Ref. [3]. This is more evident in the density of states, which vanishes at the Dirac cone, as highlighted in the zoomed-in energy range shown in the 3rd panel of Figure S3**a** for orientation A and Figure S3**b** for orientation B. The energy shift for both orientation is nearly identical which means that the charge transfer does not depend on the relative orientation. This results are comparable to the observation in Ref. [4, 5].

By unfolding the electronic band structure of the super cell, as shown in Figure S3, we clearly show the Dirac cone shift, see Figure S5**a**. The pristine Graphene bands (shown in Grey) reproduce by unfolded bands by a global band energy shifted of $\approx 0.7$ eV. This indicates that there is a weak hybridization between the graphene and $\alpha$-RuCl$_3$ layers but the charge transfer plays a significant role. In Figure S5**b** we display the unfolded $\alpha$-RuCl$_3$ band structures obtained by non-spin polarized GGA simulation. The $\alpha$-RuCl$_3$ bands do not show significant deviations from the pristine case. A zoomed-in region close to the Fermi energy shows a slight down shift of the bands, see Figure S5**c**.

## CHARGE TRANSFER REDISTRIBUTION

Using Bader [6] analysis we analyze the charge transfer depletion and accumulation for different intercalated crystals. First, we analyze the charge transfer as function of the c lattice constant in the tensile single-layer stacking (A|A), and compare it to the non-tensile single-layer stacking crystal. We find that the charge transfer is strongly affected by the graphene layer-$\alpha$-RuCl$_3$-distance $d_{\mathrm{Gr-Ru}}$. At a distance $d_{\mathrm{Gr-Ru}}$ which reassembles the distance in the unstrained structure, the charge transfer is almost identical, see Figure S4**a**. Next, we analyze the charge transfer as function of the intercalation stacking, results are summarized in Figure S4**b**. Our analysis shows that by increasing the number of the intercalated

---
* razpopov@itp.uni-frankfurt.de
† shirinm@clemson.edu
‡ tmatsuoka@up.edu.ph
§ valenti@itp.uni-frankfurt.de
¶ dmandrus@utk.edu

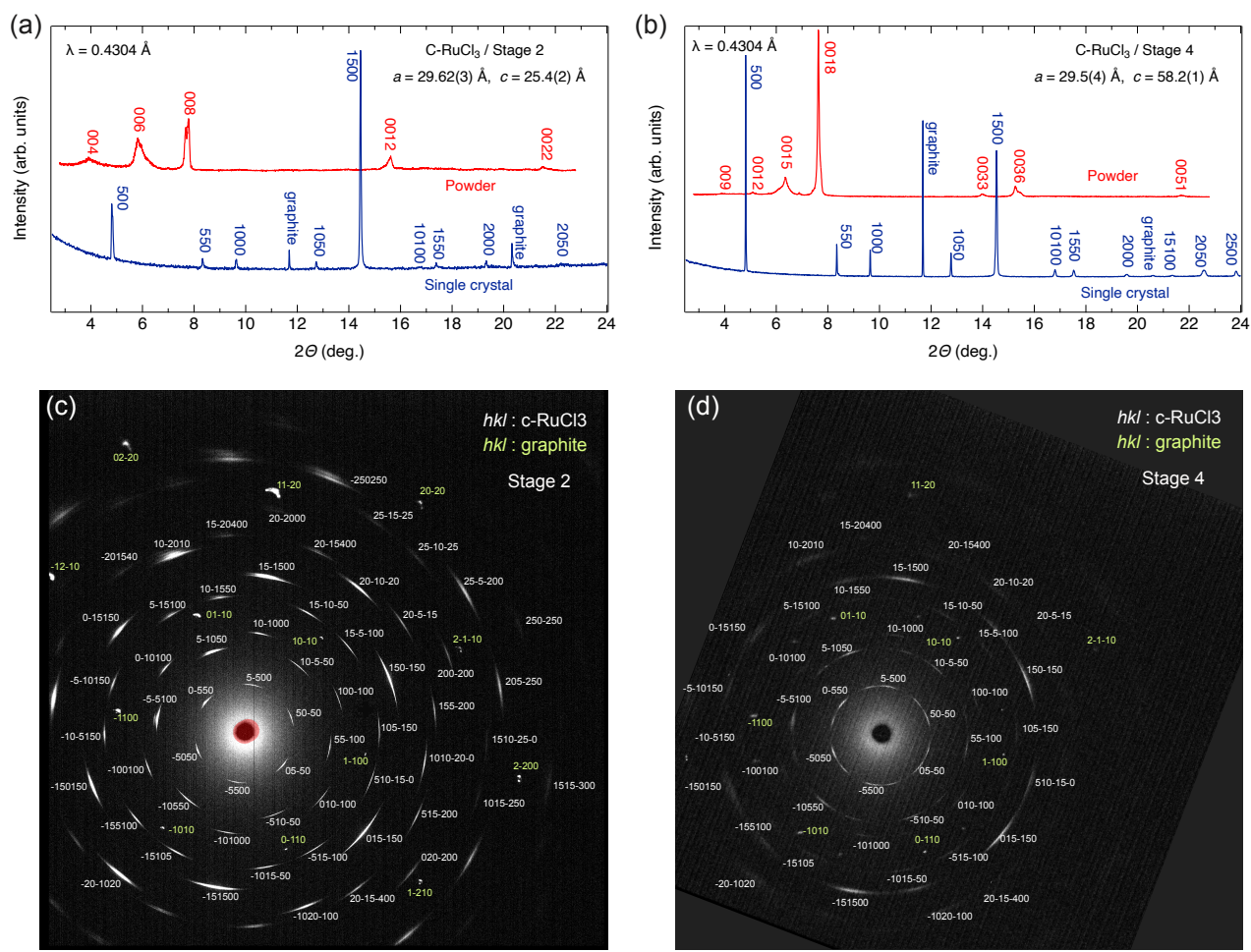

**Figure S1:** XRD images and the integrated XRD profiles for stage 2 and stage 4 C-RuCl₃ samples. **a** XRD images of the stage 2 (S2) single crystal and **b** the stage 4 (S4) single crystal samples. The *hkl* indicies are shown together. **c** The integrated XRD profiles for S2 powder and single crystal samples, and **d** for S4 powder and single crystal samples. To compare the XRD profiles between the powder and single crystal samples, the powder-XRD data obtained using $\lambda = 1.541$ Å x-ray are converted to the data for $\lambda = 0.4304$ Å. Note that the actual state of the single crystal samples is highly oriented polycrystalline, thus 'nearly single crystalline.' Due to the forward scattering geometry of the XRD instruments and the limited possible oscillation angles, the *hkl* ($l \neq 0$) reflections cannot be observed.

graphene layers the charge transfer increases significantly. A comparison of the charge accumulation in the ABAB|ABAB and ABCA|ABCA stacking configurations reveals only negligible differences, indicating that the relative stacking orientation plays a minor role in determining the charge transfer magnitude.

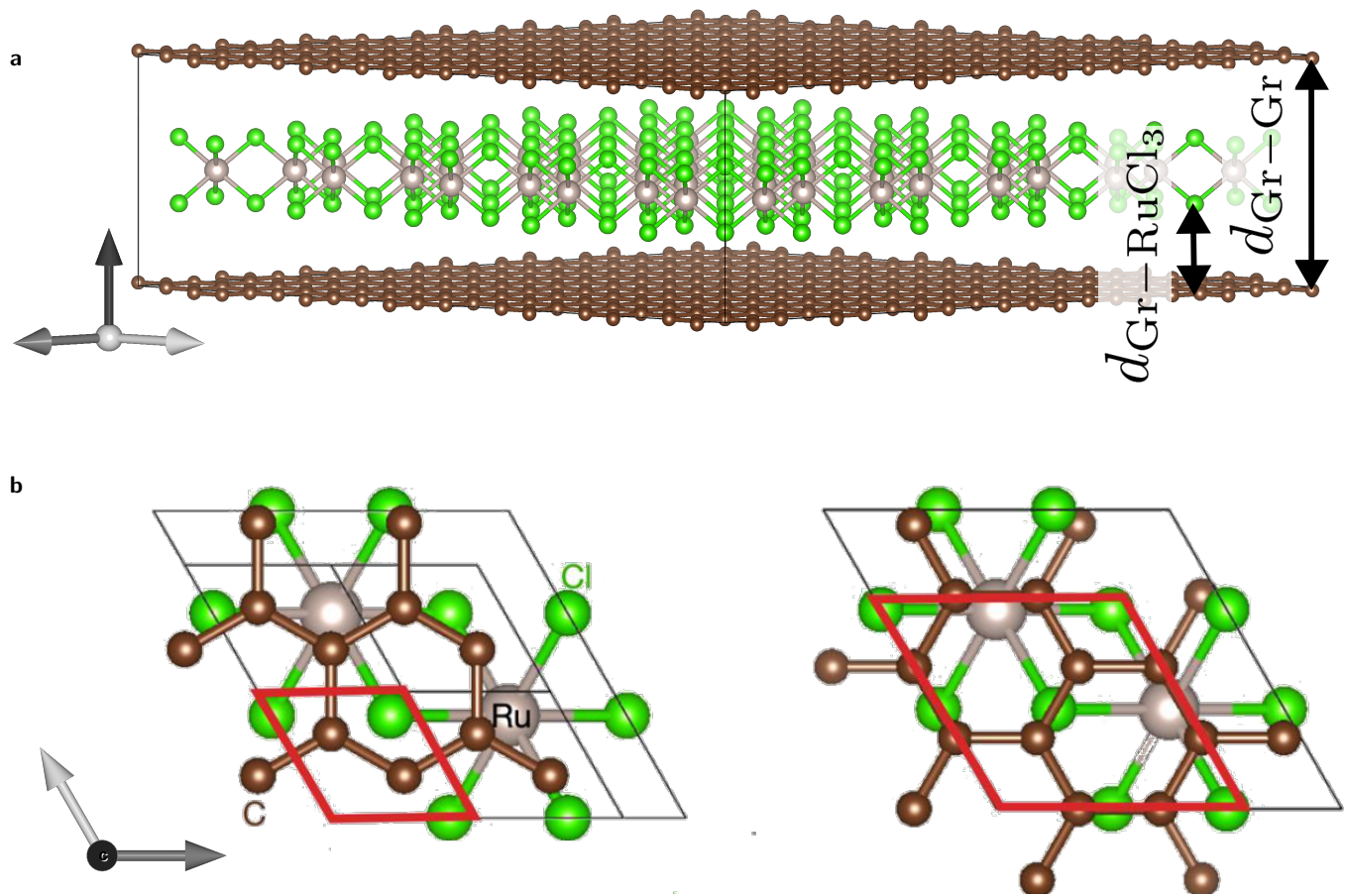

**Figure S2:** Bulk crystal structure of $\alpha$-RuCl$_3$ intercalated in graphite. **a** side view showing the alternating graphene and $\alpha$-RuCl$_3$ layers in the **c**-direction. **b** The two different relative orientations A and B considered in the study. On the left is the orientation A and on the right the orientation B structure.

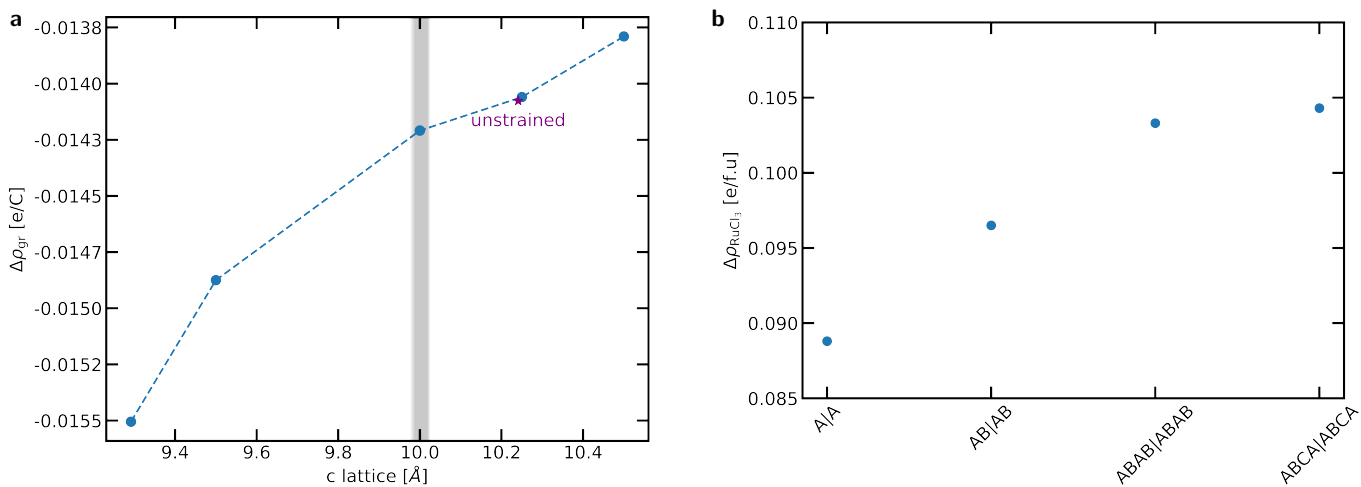

**Figure S4: a** shows the calculated charge transfer in the Gr layer as function of the c lattice constant in the tensile strained single-layer intercalation stacking (A|A). The purple data points presents the unstrained single-layer stacking structure. In the gray shaded area the $d_{\mathrm{Gr-Ru}}$ of the strained and unstrained crystals are almost identical. Please note, the dashed lines between the data points are only an eye guide. **b** Calculated charge transfer on the $\alpha$-RuCl$_3$ layer for each intercalation, A|A, AB|AB, ABAB|ABAB, and ABCA|ABCA. Positive sign refers to charge accumulation and negative to charge depletion.

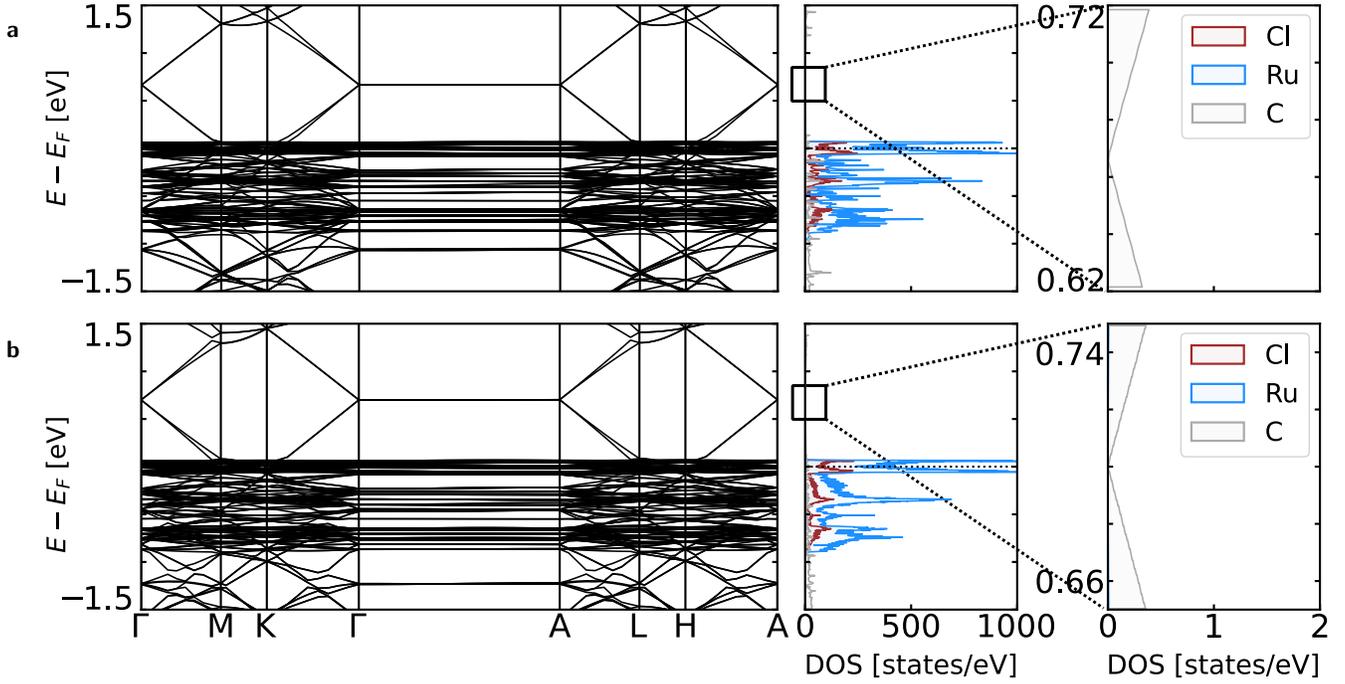

**Figure S3:** Calculated electronic band structure and atom-resolved density of states (DOS) within GGA for the relative orientations **a** A and **b** B. The 3rd panel is a blow-up of the shifted Dirac cone of graphene due to the charge transfer between graphene and $\alpha$-RuCl$_3$.

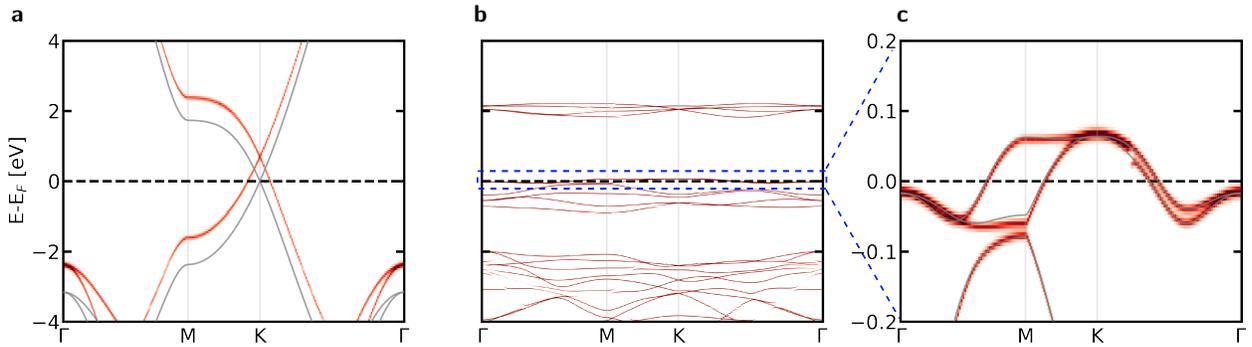

**Figure S5:** Unfolded band structure of zero-strain $\alpha$-RuCl$_3$ layer intercalated with single layer Graphene layers obtained via non-spin polarized GGA calculation. The unit cell consists of 5×5-$\alpha$-RuCl$_3$ and 12×12-Graphene. **a** shows the unfolded Graphene electronic bands. **b** shows the unfolded $\alpha$-RuCl$_3$ electronic bands, and **c** shows a zoomed energy window around the Fermi energy of the $\alpha$-RuCl$_3$ bands. The grey lines represent the electron bands of the pristine Graphene and $\alpha$-RuCl$_3$ crystal structures.


\end{document}